\documentclass[lettersize,journal]{IEEEtran}
\usepackage{amsmath,amsfonts}
\usepackage{algorithmic}
\usepackage{algorithm}
\usepackage{array}
\usepackage[caption=false,font=normalsize,labelfont=sf,textfont=sf]{subfig}
\usepackage{textcomp}
\usepackage{stfloats}
\usepackage{url}
\usepackage{verbatim}
\usepackage{graphicx}
\usepackage{cite}
\usepackage{color}

\begin{document}

\title{Enabling Augmented Segmentation and Registration in Ultrasound-Guided Spinal Surgery via Realistic Ultrasound Synthesis from Diagnostic CT Volume}

\author{
Ang Li$^\dag$,
Jiayi Han$^\dag$,
Yongjian Zhao,
Keyu Li,
Li Liu$^\ddag$
\thanks{$^\dag$ These authors contributed equally}
\thanks{$^\ddag$ represents the corresponding author. Email: lliu@ee.cuhk.edu.hk}
\thanks{Ang Li, Yongjian Zhao, Keyu Li and Li Liu is with Chinese University of Hong Kong.}
\thanks{Jiayi Han is with Fudan University.}

}


\maketitle

\begin{abstract}
This paper aims to tackle the issues on unavailable or insufficient clinical ultrasound (US) data and meaningful annotation to enable bone segmentation and registration for US-guided spinal surgery. While the US is not a standard paradigm for spinal surgery, the scarcity of intra-operative clinical US data is an insurmountable bottleneck in training a neural network. Moreover, due to the characteristics of US imaging, it is difficult to clearly annotate bone surfaces which causes the trained neural network missing its attention to the details. Hence, we propose an \textit{In silico} bone US simulation framework that synthesizes realistic US images from diagnostic CT volume. Afterward, using these simulated bone US we train a lightweight vision transformer model that can achieve accurate and on-the-fly bone segmentation for spinal sonography. In the validation experiments, the realistic US simulation was conducted by deriving from diagnostic spinal CT volume to facilitate a radiation-free US-guided pedicle screw placement procedure. When it is employed for training bone segmentation task, the Chamfer distance achieves 0.599mm; when it is applied for CT-US registration, the associated bone segmentation accuracy achieves 0.93 in Dice, and the registration accuracy based on the segmented point cloud is 0.13$\sim$3.37mm in a complication-free manner. While bone US images exhibit strong echoes at the medium interface, it may enable the model indistinguishable between thin interfaces and bone surfaces by simply relying on small neighborhood information. To overcome these shortcomings, we propose to utilize a Long-range Contrast Learning Module (LCLM) to fully explore the Long-range Contrast between the candidates and their surrounding pixels. In the ablation experiments, it is verified that the proposed Long-range Contrast Learning module is effective in the precise positioning of the US region of interest. On top of that, the training data is entirely generated by our proposed US simulation framework without fine-tuning based on real clinical data, which demonstrates its effectiveness of the bone realistic US simulation framework.

\end{abstract}

\def\abstractname{Note to Practitioners}
\begin{abstract}
The motivation of this paper is to address the issues on unavailable or insufficient bone US images and annotation labels. We employ such a data augmentation technique to generate realistic simulated bone US and annotation associated with the corresponding CT volume.
The problems of current US data augmentation approaches are mainly the inability to generate continuous context-accurate data (neural network-based algorithm) and the lack of real-time physical simulation capability (traditional acoustics-based algorithm).

In this paper, we enhance the US simulation derived from CT images by supplementing techniques such as extrusion simulation, moving space simulation, and image augmentation to yield higher quality images, and hence apply these images directly to the target task, i.e., train a neural network to guide US-guided spinal surgery.
Besides, we demonstrate the effectiveness of the realistic US simulation framework and each new module of the neural network using ablation experiments. It can be concluded that the neural network trained with the data generated by US simulation framework promises to enable US-guided pediclescrew placement procedures.
In the future work, we will apply the \textit{In silico} bone US simulation as a reinforcement learning environment and deploy the trained agents to directly guide US-guided procedures.
    
\end{abstract}

\section{Introduction}
\label{sec:intro}

Segmentation of bone surfaces from intra-operative US data followed by CT-US registration is two critical steps for US-guided spinal surgery. Recent research has focused on the use of deep learning-enabled methods for accurate, robust, and real-time segmentation and registration of bone surfaces. However, scarcity of data size, due to a lack of standardized data and patient privacy concerns, is a major challenge in applying deep learning-enabled methods in the intra-operative imaging ﬁeld. This is speciﬁcally a challenge due to the fact that the US is not a standard imaging modality in spine-related surgeries and US-guided spinal surgeries are not common; even if spinal sonography can be available for pre-clinical collection procedures, annotations would be still a severe challenge due to the vast amounts of data. Another limiting factor is the manual data acquisition process: sub-optimal orientation of the US transducer with respect to the imaged spinal anatomy will result in the user-dependent acquisition of low-quality bone scans.

Increasing the size of existing datasets through data augmentation in order to improve models’ performance is extensively investigated. Among them, a fundamental approach to obtaining a large labeled dataset is \textit{In silico} realistic simulation of spinal US images. The US simulation methods could be broadly categorized into three types: acoustic model-based US simulation, image-based US simulation, and generation of a virtual image using a generative adversarial network (GAN). US generation algorithms based on acoustic models are usually very slow \cite{wang2012modelling}. Generating US by GAN may require training different networks for different organs, meanwhile, obtaining corresponding US volume and CT volume pairs aligned in the same space is difficult, allowing for the generation of large-scale synthetic US images using GAN a challenging task. Image-based approaches attempt to utilize a simulated US probe to re-sample the original image  (such as a CT scan) and then consider the image scalar as acoustic parameters of organs and then simulate propagation with acoustic properties. Nonetheless, different from other imaging methods such as CT and MRI, the US is significantly influenced by gas. To avoid its influence, the operator presses the probe to remove the gas between the probe and the skin, which leads to the distortion of tissues in the original images. However, this distortion is ignored by the former researchers. Moreover, because each tissue has a different reflection rate, the former works segment the image and design specific transfer functions to each tissue, which is time-consuming. In addition, the conventional use of US image generation is only to assist in 2d-3d image registration \cite{ning2021autonomic} or for training examining physicians \cite{piorkowski2012transesophageal} without further application of synthetic US images.

\begin{figure*}[htbp]
    \centering
    \includegraphics[width=.8\textwidth]{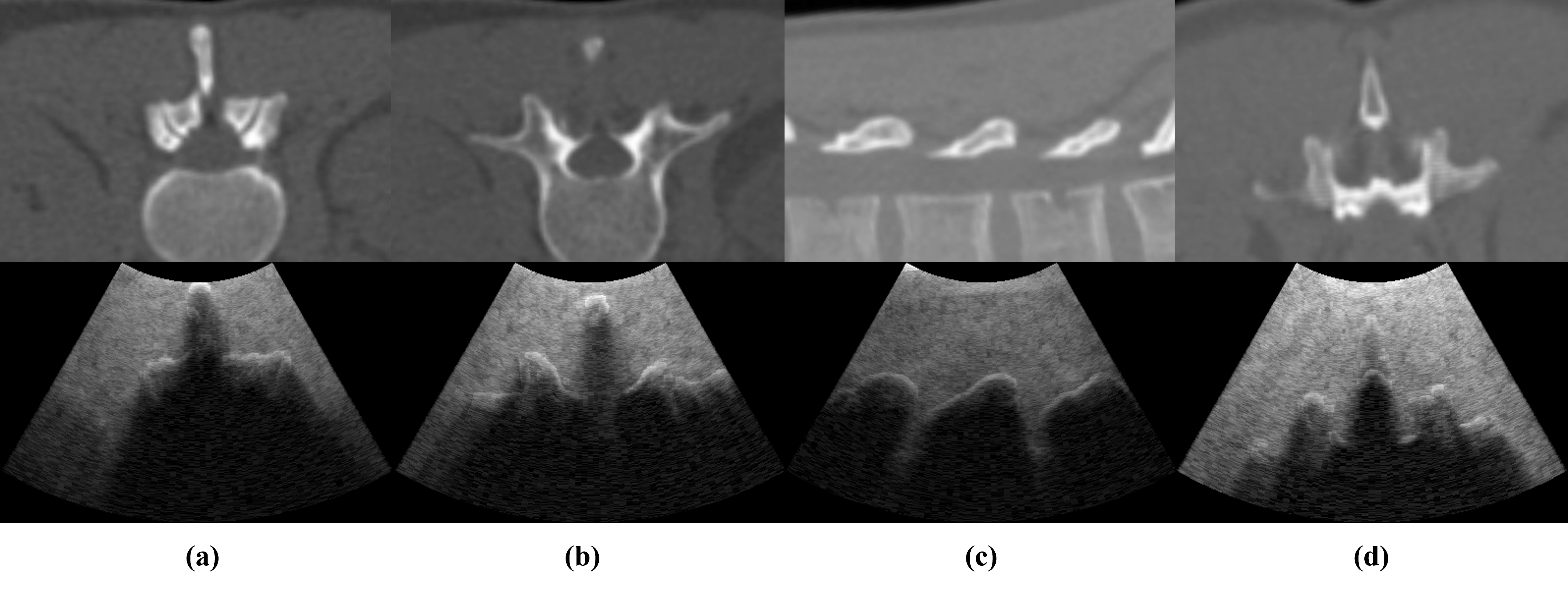}
    \caption{The figure shows examples of generated Ultrasound images which are the junctional surfaces of the two vertebrae, the plane where the vertebral plate could be entirely verified, the plane swept along the spinous, and an arbitrary scan plane, respectively.}
    \label{fig:us_example}
\end{figure*}

In this paper, we propose a novel CT-derived realistic US synthesis framework incorporating automated image generation with sampling methods, as shown in Fig.~\ref{fig:us_img_pressed}. From (a)$\sim$(b),  each column represents the real US, corresponding CT scan, reflection map, and transmission map, respectively. We simulate the distortion resulting by pressing the probe via warping the original image and propose an adaptive transfer function that could be directly adopted to the whole image which eliminates the transfer function designing process for each tissue and highly speeds up the US simulation task. To fully take the advantage of the proposed simulation and conduct real-time US image segmentation, we further propose a lightweight vision transformer with Long-range Contrast Learning Module (LCLM) which utilizes a designed cascaded dilated convolution layers to achieve dense super-large receptive field which enhances the US image segmentation. Experiments demonstrate the proposed simulation system achieves state-of-the-art performance compared with other approaches and benefits the proposed vision transformer for real US image segmentation.

Our contributions are listed as follows:
\begin{enumerate}
    \item We propose an \textit{In silico} bone US simulation framework that synthesizes realistic US images from diagnostic CT  volume.
    \item We develop a lightweight vision transformer model that achieves precise and real-time bone segmentation for spinal sonography images.
    \item Experiments demonstrate that the proposed \textit{In silico} bone US simulation approach dramatically enhances the segmentation performance in comparison with initial CT scans, indicating that the proposed data augmentation method is capable of pre-training models for real clinical spinal sonography.
\end{enumerate}
\section{Related Work}
\label{sec:rel}

In this section, the related work on realistic US simulation and associated bone segmentation are discussed.

\subsection{Realistic US Simulation}

Traditional acoustics-based US simulation software was pioneered by Jenson et al. \cite{jensen1996field} in 1996.
 K-space method for fast computation of pulsed photo-acoustic fields was proposed by B. T. Cox et al.\cite{cox2005fast} in 2005 and Treeby et al.\cite{treeby2012modeling} optimized the US image propagation model on the K-space method and then the widely used K-wave tools have been developed. The advantage of this conventional acoustic-based algorithm is that it can simulate various types of US imaging systems with the physical effect of the images as realistic as possible.
 
As these algorithms based on physical models of acoustics have efficiency problems to generate large datasets, researchers started to develop ray-tracing-based approaches. Burger et al. \cite{burger2008simulation} developed a US simulation system by segmenting the CT dataset into different tissues, and then assigning velocity, impedance, scattering factor, and other acoustic properties to each tissue; hence, it was used to simulate the sound propagation, absorption, and scattering processes. Cong et al. \cite{ cong2013fast} proposed a multi-scale enhancement method to augment tubular structures to simulate blood flow, and allow for US images more realistic. 
Piorkowski et al. \cite{piorkowski2012transesophageal} in 2013 applied the algorithms of Wein \cite{wein2008automatic} and Kutter\cite{kutter2009visualization} to make a Transesophageal Echocardiography (TEE) Simulator which has a tremendously positive effect on training doctors to perform TEE examinations. To further accelerate the image generation speed, Wang et al.\cite{wang2020real} used NVIDIA's Optix 6.0 ray-tracing engine to do the Monte Carlo simulation of US with a good result compared to GAN and Field-II \cite{jensen2004simulation}.

In recent years, generative adversarial network (GAN) models have been used extensively in realistic US simulation research. It generates realistic US images after learning from a large dataset in the US domain. Hu et al. \cite{ hu2017freehand} applied a GAN model to yield US images for Freehand scanning, the model takes the spatial location information as a conditional input, and then outputs the US image at the current location. This work contributes to producing US data for the corresponding location, yet they were unable to create synthetic US images for each specific patient. To address this shortcoming in 2018, Tom et al.\cite{tom2018simulating} employed cascaded GAN with an image segmentation label as conditional input to create more realistic US images and to be able to produce synthetic US images for different segmentation results. Nonetheless, according to their report, it can be concluded that even with precise segmentation the GAN-based virtual US system still has difficulty in ensuring the edge intensity and shape as the real US in same.

\subsection{US Segmentation}

Some early study utilized handcrafted features for US segmentation, such as active contour \cite{liu2010probability, talebi2011medical}. In recent years, the most popular model in US segmentation is UNet \cite{ronneberger2015u}. Many works adopt different reinforcements in UNet-based models for US segmentation. \cite{zhang2021sha} proposed a multi-task UNet which combines classification and segmentation tasks. \cite{shuvo2021cnl} proposed a lightweight UNet model which alternately adopt $3\times 3$ and $1\times 1$ convolution layers. They also introduced a false output suppression mechanism that combines patch-wise classification and segmentation results to eliminate false positive. \cite{pan2021sgunet} adopts spatial attention on US image segmentation task.
\section{Method}
\label{sec:method}

\subsection{Realistic US Simulation from CT Volume}
The entire US simulation progress is divided into several parts: data sampling, data filtering, US transmitting, and image blending.
\paragraph{Image Probe and Press simulation}
Since the goal of the proposed US simulation system is to train doctors or artificial intelligence agents for surgical robots, the moving space of probe motion cannot be the entire 3D space. The probe in the system can move in the provided scanning space, but the probe is a curved surface, it is the probe surface, and the scanning space will not be fitted perfectly.  The scanning space will wrap around and fit the probe if the pressure is simulated. However, the CT volume will not deform with the surface. Hence, an algorithm needs to be proposed to make the CT data deform with the probe surface along the scanning space. Using the spring-mass model to simulate deformations in scanning space by using the moving least squares algorithm to fit the CT deformation is certainly good, but also has two demerits. The first is solving the spring-mass simulation and the moving least squares equation will consume a lot of computational resources. Secondly, the algorithm requires a lot of extra work to mark the anchor points on the images. Under this situation, we changed the local translation image warping algorithm proposed by A. Gustafsson for CT data and apply it with the shape of the probe to the image. The algorithm only needs to know the shape of the probe and the HU value of the CT data of that slice to simulate the motion of the tissues. If further acceleration is desired, the weight value of the HU parameter can be set to 1, then the UV coordinate mapping only needs to be calculated once, without affecting the imaging efficiency at all.

The equation is formulated as Equ. \ref{equ:press_warp}

\begin{equation}
\left \{
    \begin{array}{l}
         \vec{u} = \vec{x} \frac{r_{max}^2 - |\vec{x} - \vec{c}|^2}{r_{max}^2 - |\vec{x} - \vec{c}|^2 + D}(\vec{m} - \vec{c})\\
    \\
    D = \frac{100}{f}\alpha(hu)|\vec{m}-\vec{c}|^2
    \end{array}.
\right.
\label{equ:press_warp}
\end{equation}

\begin{figure*}
   \centering
   \includegraphics[width=0.8\textwidth]{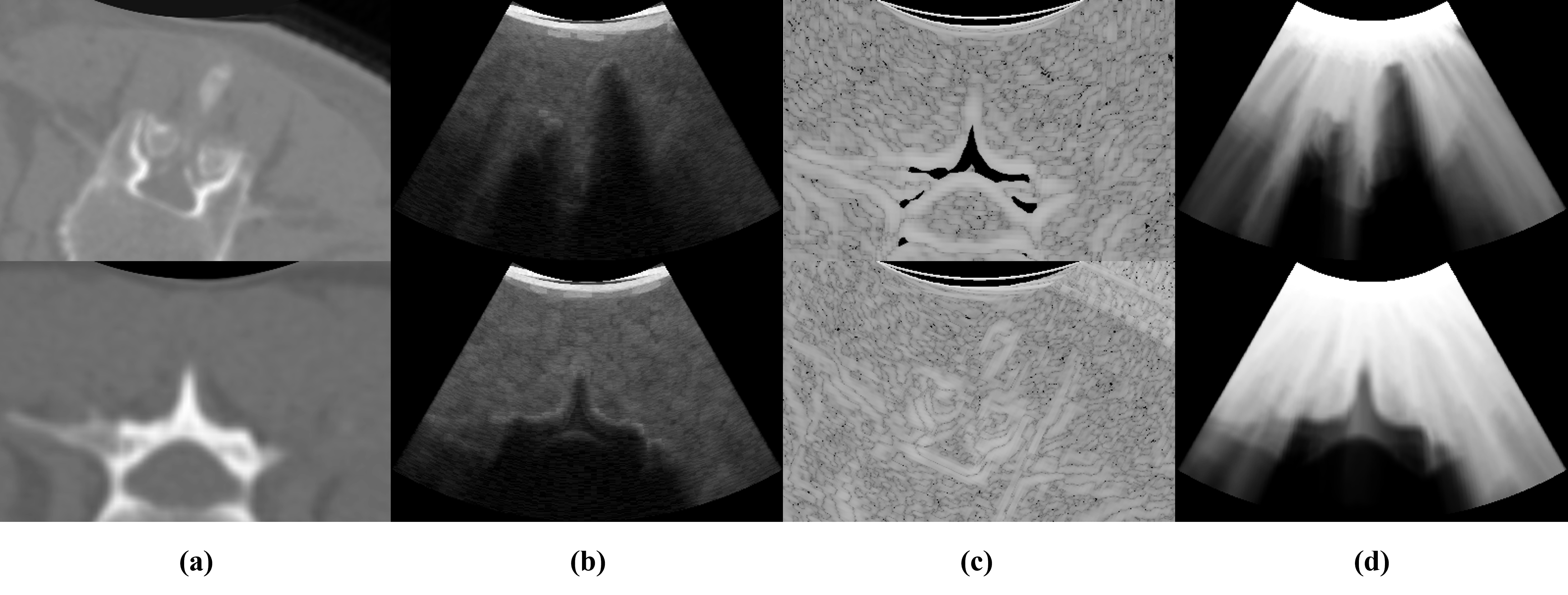}
   \caption{Figure (a) shows the CT images after being processed by the pressure simulation, the skin and muscles are compressed into the probe's curved surface. Figure (b) shows the virtual US image produced with simulated extrusion. Figure (c) shows the reflection map after extrusion. Figure (d) shows that the processed image will not affect the propagation.}
   \label{fig:us_img_pressed}
\end{figure*}

  In the above equation, $f$ controls the ratio of deformation. In our case, we want to push the intersection point of the center-line of the probe and the top line of the sampled image, as the blue dot is shown in Fig.\ref{fig:press_range}, to the top of the probe which is shown in the figure as the red dot. According to the mentioned algorithm, all the tissue between the blue and red dots needs to be pressed downwards. Assuming that the tissue is a rigid body, the thickness of the tissue does not change by pressing with a very big force. In this case, the original position at the red dot should be pushed to the top of the green curve. 
Then a green arc that represents the limit of tissue movement can be drawn. The squeezing of the tissue in the area between the probe arc and the green arc will cause high reflections or absorption in this area. Hence, in the process of simulating transmission, it is significant to make the sound waves attenuate less in this area, otherwise, the squeezed tissue generated by the algorithm will completely block the transmission of US. As you can see from Fig.~\ref{fig:us_img_pressed}, there are a few bright lines below the probe curve, which are the signal of the squeezed tissue and it is identical to the vivo US image.

\begin{figure}[htbp]
    \centering
    \includegraphics[width=0.4\textwidth]{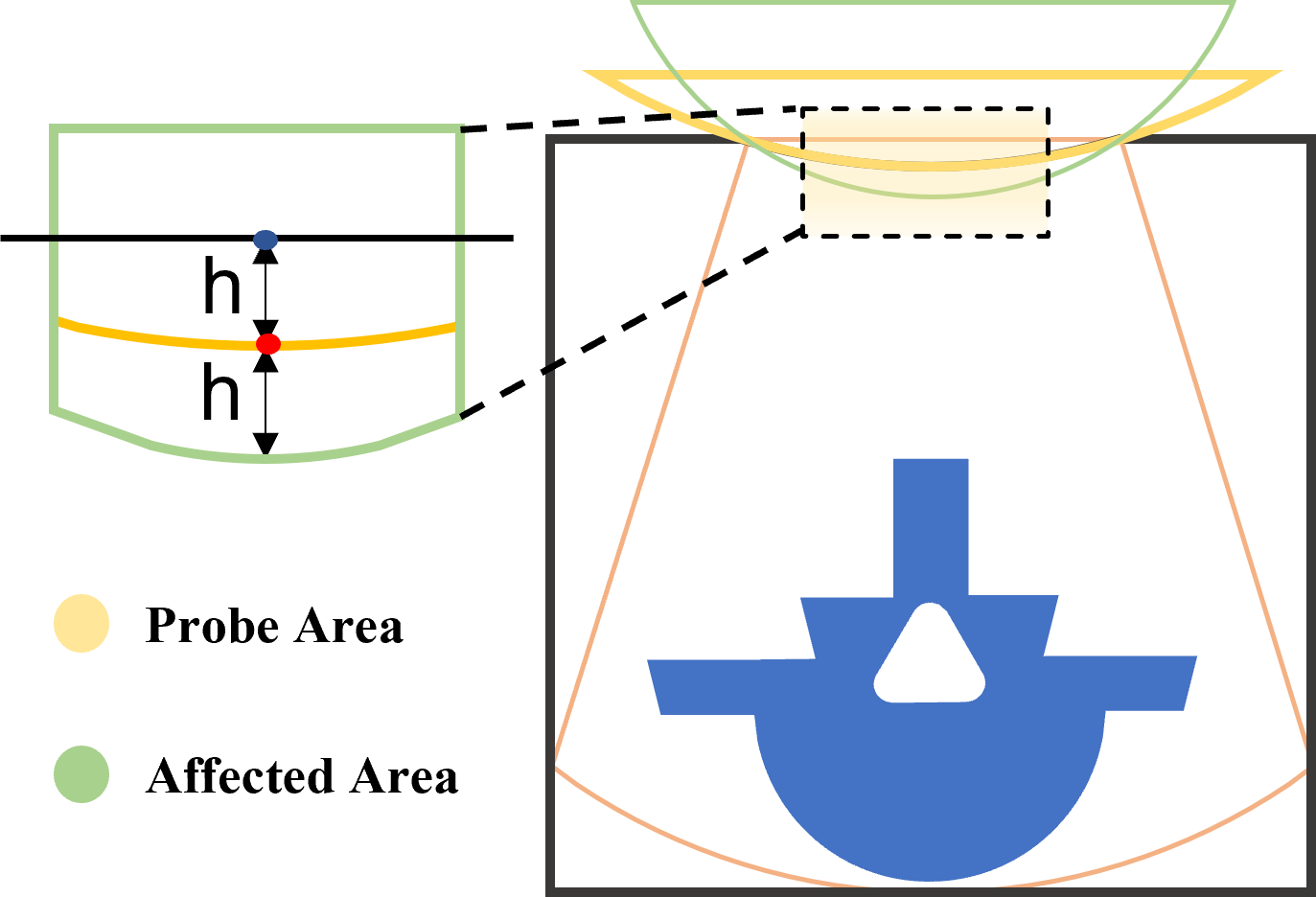}
    \caption{This figure illustrates the pressure simulation algorithm. As an enlarged part of the dashed box in the figure, the yellow arc represents the shape of the probe, and the area between the yellow arc and the green arc is the target space being squeezed into.}
    \label{fig:press_range}
\end{figure}

\paragraph{Restricted Movement Space of Probe}

As mentioned above, the probe cannot move in space arbitrarily. To automate the acquired US images, two types of probe moving regulation are designed in this paper using spinal US as an instance.
Define $M$ as a mesh in 3d space, clipping a 3D sub-mesh $M_r$ of interest by the bounding box of the scanning target, the movement of the probe is only meaningful in this manifold. The first way is to move freely on the grid, for any point $p$ on $M_r$, the direction of movement of p will be curved along the grid normal vector. In this case, a very small $dx$ or $dy$ will not make a great $dz$, which would not make the image change dramatically and could not acquire a continuous image. 
The second method is to slice $M_r$ into a series of curves in 3D space along the forward direction $M_{r0}... M_{rn}$. In each polyline, the points can only move forward or backward along the tangent direction by a certain distance. It has the merit that keeping the components of the probe's normal vector remain consistent across that curve, which is often used for automatic circular or flat scanning of a target and ensuring that the target being swept is on a line or a point. 
Besides, both of these movement methods can be employed by a reinforcement learning-based agent, and the first method is more suitable for continuous control, while the second method is more suitable for discrete control.

\begin{figure*}[htbp]
    \centering
    \includegraphics[width=0.8\textwidth]{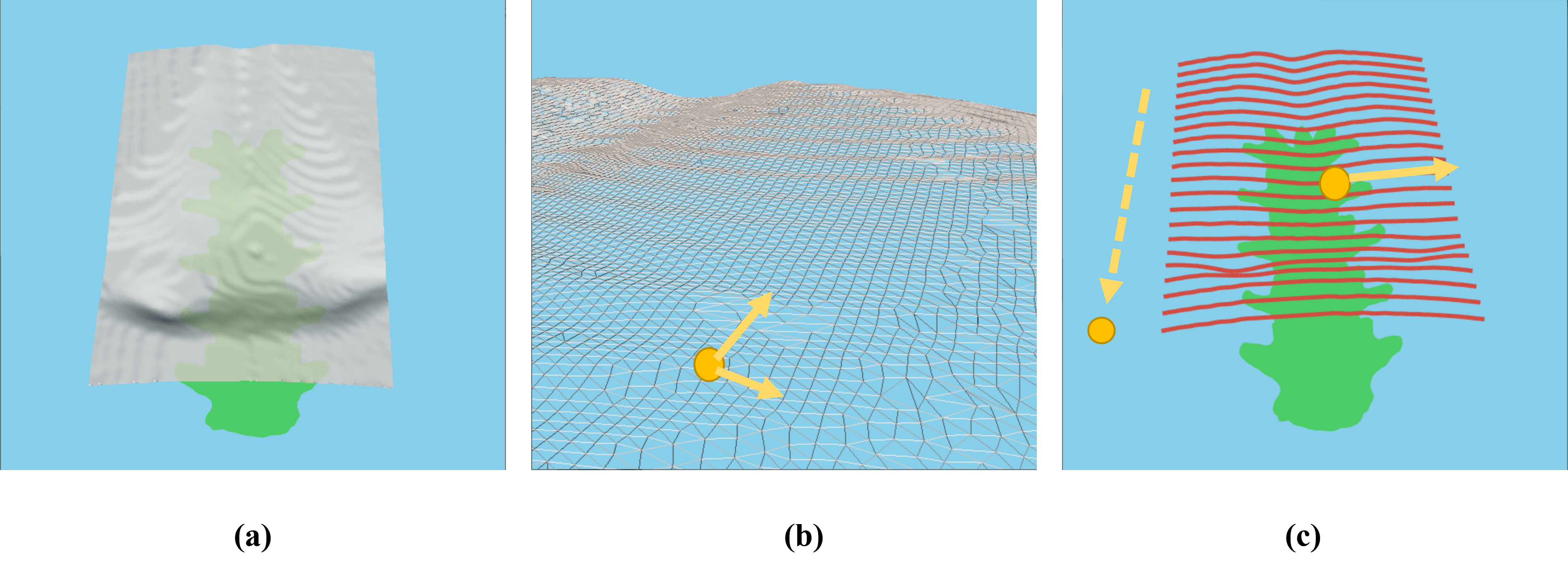}
    \caption{The green part of the figure is the object to be scanned. Figure (a) shows the relationship between the entire feasible space (skin) and the scanned object(spine). Figure (b) shows the first type of movement, with the yellow dot representing the tip of the probe and the arrow representing the two-degree freedom of movement in the manifold. Figure (c) shows the second type of movement, the red line represents the discrete surface, the dashed arrow represents the probe can switch on different curves, and the arrow represents the probe can be moved along the curve direction.}
    \label{fig:press_range}
\end{figure*}

\paragraph{Sound Reflection}

The first step in the simulation is to map the CT-HU image into an acoustic impedance image, with the mapping look-up table references to \cite{raum2004frequency}. In the mapping table, each HU value will be mapped to an acoustic impedance Z. The image of this acoustic impedance will then be processed instead of the CT image.
Based on the acoustic physical transfer properties, we first introduce the Fresnel equation to calculate the reflection. Dividing sound waves into those parallel and those perpendicular to the plane, the equations are:
\begin{equation}
\left \{
    \begin{array}{l}
        R_\perp=(\frac{Z_1cos\theta_i - Z_2cos\theta_t}{Z_1cos\theta_i + Z_2cos\theta_t})^2
    \\
    R_\parallel=(\frac{Z_2cos\theta_i - Z_1cos\theta_t}{Z_2cos\theta_i + Z_1cos\theta_t})^2
    \\
    R = \frac{1}{2}(R_\perp + R_\parallel)
    \end{array}.
\right.
\label{equ:fresnel_equation}
\end{equation}

When sound waves transmitted from one medium to another, it is not only be reflected but also be refracted.
This refracted has the same physical properties as the refracted of light passing through different media which is given by Snell's law:

\begin{equation}
\frac{sin\theta_1}{sin\theta_2} = \frac{Z_2}{Z_1}
\label{equ:snell_law}
\end{equation}

In fact, when a sound wave is refracted or bent, the direction will be kept until the next transmission. It is possible to simulate multi-level refraction and reflection values using a recursive ray-tracing algorithm
However, since the amplitude of the refracted sound waves is small and has little effect on the generated image quality, the proposed method does not consider its direction after refraction and only superimposes the amplitude on the incident sound wave of the next medium.

Irradiance: 
Lambert's cosine theorem is introduced to take the irradiance of the sound wave should into consideration. This equation means that the more perpendicular the direction of the tissue gradient and the sound wave, the stronger the sound reflection and the brighter the tissue.
\begin{equation}
\frac{I_r}{I_i} = cos\theta
\label{equ:cosine_law}
\end{equation}
\paragraph{Sound Attenuation}

Attenuation in the propagation of US passing through tissue could be broadly divided into reflection, scattering, and absorption, where absorption accounts for the majority of it.
The paper \cite{4189}states that US absorption of substances such as bovine and porcine livers accounts for $90\%$ or even $100\%$ of US propagation attenuation.
Even in the tissues that exhibit anisotropy such as the bovine brain and leg muscle absorption still makes a major contribution to the attenuation.
Therefore, mapping the CT scan to an absorption image is important.
The mapping table is obtained by interpolating the data\ref{tab:absportion} in the paper \cite{hoskins2007physical}.
The formula for absorption is given by the Lambert-Beer law. Since different tissues may not have the same US absorption capacity in the same Hu value, in order to get a better result from the generated image, an adjustable parameter $\alpha$ is added to the original equation. The modified equation is as follows:

\begin{equation}
I_a = I_010^{-\alpha*d*f*\frac{1}{10*\beta}}
\label{equ:Lambert_Beer_law}
\end{equation}

\begin{table}[htbp]
    \centering
    \begin{tabular}{c|c|c|c|c}
    \hline
         Tissues   & Density     & Velocity   & Impedance    &Attenuation \\
    \hline
         Skin  &   1100 &      1631 &  1.794&     0.22 \\
         Fat    & 916   & 1435  & 1.352     & 0.975  \\
         Muscle           & 1041       & 1595    & 1.647     & 1.47  \\

    \hline
    \end{tabular}
    \vspace{1mm}
    \caption{Acoustics properties of different tissues. with Density(kg $m^-3$), Velocity($ms^-1$), Impedance $10^6(kg m^-2s^-1)$, Attenuation($dB cm^-1MHz^-n$) \cite{hoskins2007physical}}
    \label{tab:absportion}
\end{table}

\paragraph{Propagation in Discrete Field}

\begin{figure}[htbp]
    \centering
    \includegraphics[width=0.25\textwidth]{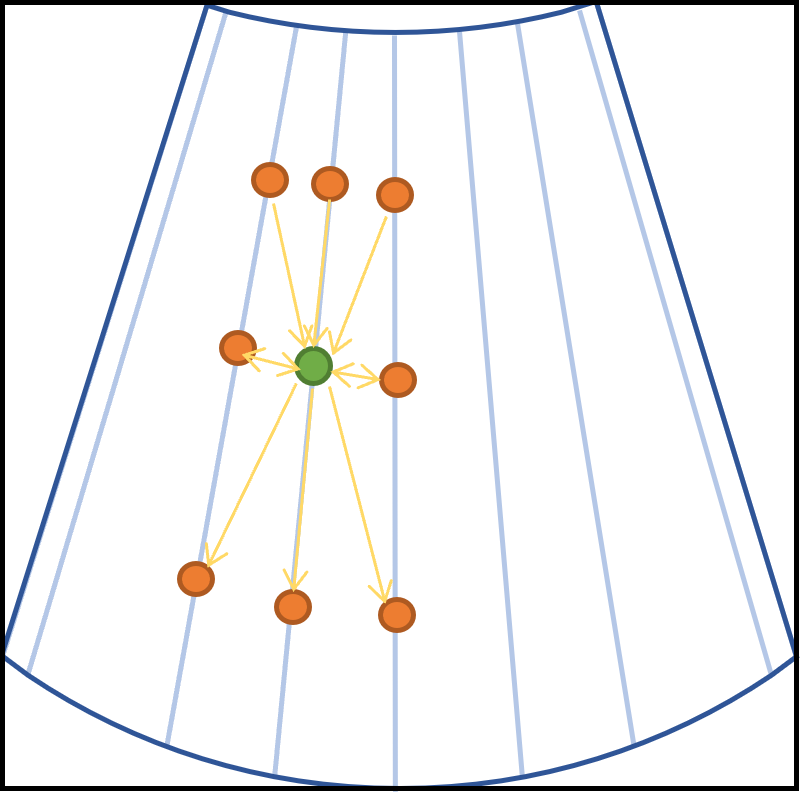}
    \caption{Illustration of propagation.}
    \label{fig:trans_illustration}
\end{figure}

The propagation of the US is the most time-consuming part of the generation process. Particularly, the steps of finding receiver tissues of reflection and refraction would take a lot of time in detecting intersections that occur frequently.

At the same time, the direction of the adjacent pixels is discrete, whereas the direction of acoustic transmission is continuous. So it may happen that the line segments intersect while the pixels do not have an intersection.  Each pixel among the eight neighborhoods should be regarded as the obstacle of the wave propagation and the normalized result of the cosine value of the direction of propagation and the direction of the pixel connection is used as the weight of the obstacle.
That is, if the connection direction is in the opposite direction to the direction of propagation, the weights will be zero and the obstacle weight is maximized when the directions are exactly the same.

In order to calculate the amplitude enhancement due to reflection and refraction easily, each point also calculates its wave propagation to the neighborhood. In this case, a negative cosine value of the pixel connection direction and the propagation direction will consider as reflection. Cosine values less than 0.8 (a hyper-parameter) will be considered as contrast enhancement by refraction and scattering. The schematic diagram of propagation is shown in \ref{fig:us_example}

\paragraph{Reflection Enhancement}

\begin{figure*}
    \centering
    \includegraphics[width=.8\textwidth]{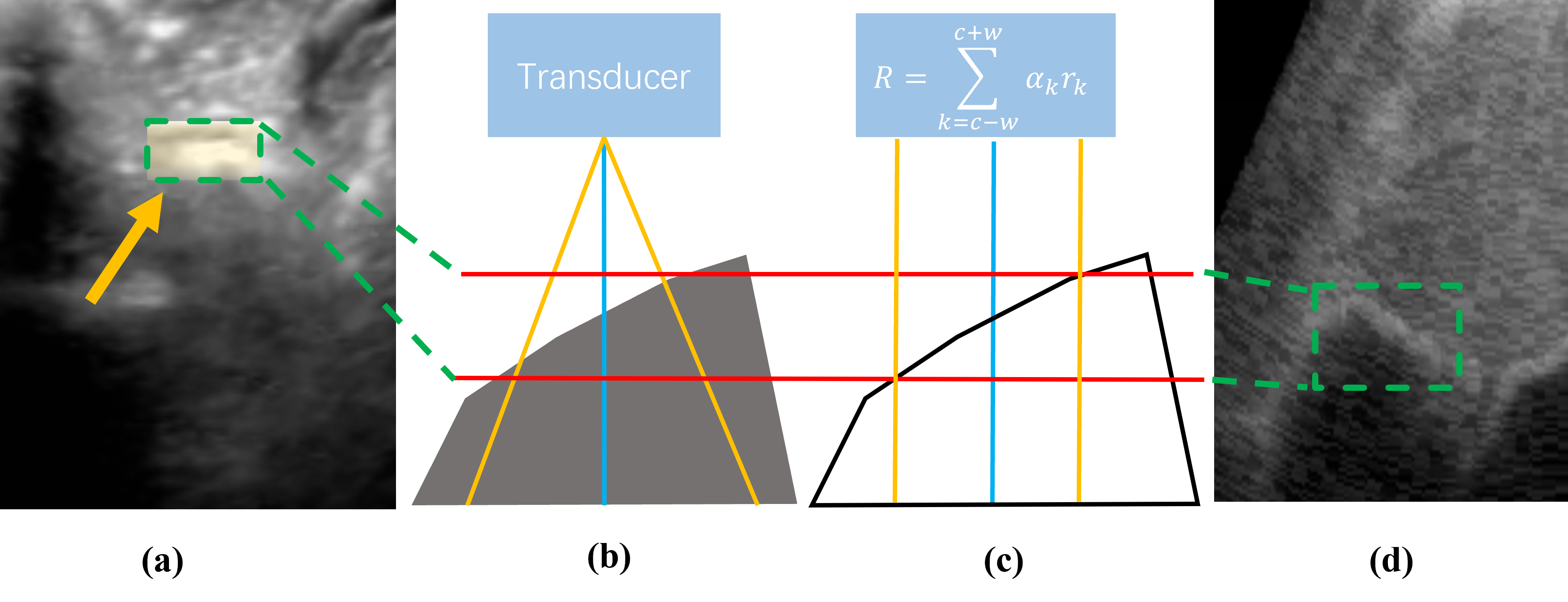}
    \caption{The thick stripe is indicated by the arrow in Figure a). Figure b) shows how the thickness of the US waves induces the thick stripes artifact. Figure c) is a schematic of sampling in a 3d gradient texture. The Blue line indicates the imaging plane. The reflection value between the blue line and the yellow line is multiplied and accumulated into the imaging plane. Figure d) shows the simulated thick stripes in the system.}
    \label{fig:thick_stripe}
\end{figure*}

When the transducer receives the reflected sound waves from the medium gap, a bright stripe appears in the US image. Because the reflection only occurs when the properties of the medium change, there should be only one reflection between the bone surface and the muscle tissue, but the US image often shows a very thick bright stripe between bone and muscles.
hacihaliloglu et al. \cite{hacihaliloglu2017ultrasound} propose that this bright stripe is produced by the thickness of the US in the elevational direction. Based on this theory, this paper sampled 3d gradient textures in front of and behind the current plane (As the blue line in figure \ref{fig:thick_stripe}). These textures are multiplied by a weight value $\alpha$ which is inversely proportional to the distance and then accumulated into the current generated US image. Simulating thick stripes is significant for neural network training; if the thick stripes are not present in the training data, the output of the model will deviate from the ground truth when segmenting the bone surface on the real US data.

In summary, the generated US image is obtained by calculating the propagation image with the reflection image and the absorption image and then blending these three images together with the radial noise.

\begin{figure*}
    \centering
    \includegraphics[width=.8\textwidth]{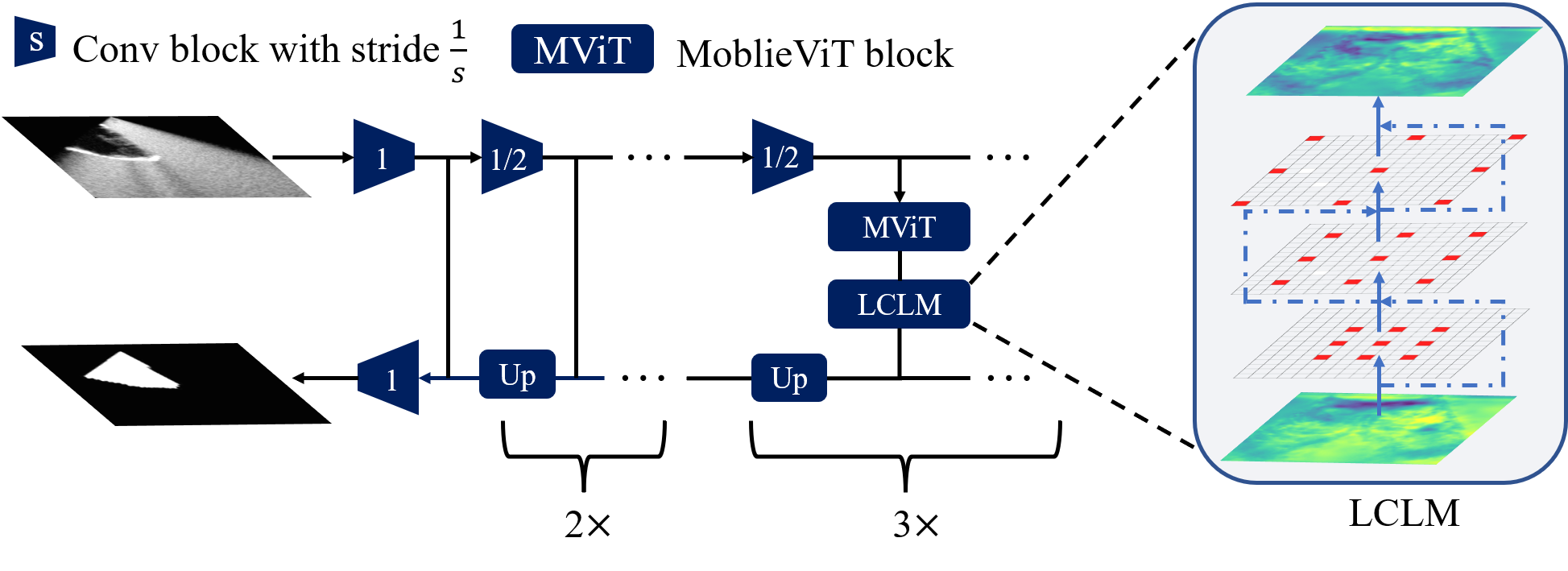}
    \caption{The framework of the proposed lightweight vision transformer for US segmentation. For the first two stages, we only utilize basic convolution layers to learn local representations. In the last three stages, we implement MobileViT block and LCLM alternately to cover both long-range dependency and long-range contrast.}
    \label{fig:framework}
\end{figure*}

\subsection{US Image Segmentation}

In this work, we propose a lightweight segmentation framework for US segmentation, as shown in Fig.~\ref{fig:framework}. We follow \cite{mehta2021mobilevit} as the backbone. After each MobileViT block, we add an LCLM for further long-range contrast learning. We then utilize a feature pyramid network (FPN) to recover the initial resolution of the input US image.

\paragraph{\bf MobileViT block} To cover the global information, MobileViT block first utilizes a $3\times 3$ convolution to aggregate the neighboring features. The pixels are then grouped into patches, each patch contains $h\times w$ pixels in which $h$ and $w$ represent the height and the width of the patch. The $i_{th}$ token of each patch transverses its feature from each other by SA. When $h<=3 \wedge w<=3$, each pixel can access information from all pixels. Details of the MobileViT block could be found in \cite{mehta2021mobilevit}.

\paragraph{\bf Why is LCLM needed}
We find that the region of interest (RoI) in US images has similar textures, but is weak in amplitude to other tissues. Basic $3\times 3$ convolutions are capable of modeling the neighboring texture but fail to learn long-range textures. As the local texture of the to-be-segmented region is similar to the other tissues, we have to design a module that covers the texture within and outside the to-be-segmented region.

\begin{figure}[htbp]
    \centering
    \includegraphics[width=0.48\textwidth]{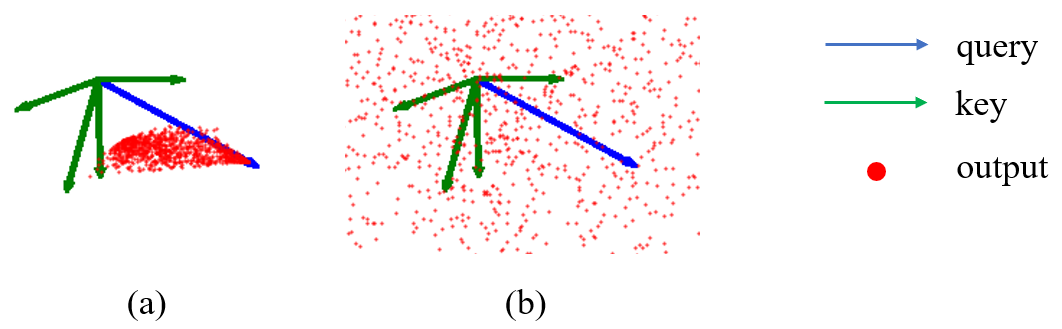}
    \caption{The possible outputs (dots in red) of self-attention (a) and convolution (b) with fixed $W_v$.}
    \label{fig:domain}
\end{figure}

In vision transformers, long-range texture modeling is achieved by self-attention (SA) of all tokens as in Equ.~\ref{equ:sa}:
\begin{equation}
    \begin{array}{l}
         \widetilde{x} = \operatorname{softmax}(\frac{xW_q(xW_k)^T}{d})(xW_v),
    \end{array}
    \label{equ:sa}
\end{equation} in which $x\in \mathbb{R}^{N\times d}$ represents the $N$ input tokens with dimension $d$, $W_q\in \mathbb{R}^{d\times d_q}$, $W_k\in \mathbb{R}^{d\times d_q}$, and $W_v\in \mathbb{R}^{d\times d_v}$ are learnable parameters. We rewrite the self-attention to the formula as in Equ.~\ref{equ:sa2}:
\begin{equation}
\left \{
    \begin{array}{l}
         \widetilde{x}_i = \sum\limits_{j\in \mathcal{A_i}}w_{j}(x_{j}W_v)\\
         w_{j} = \frac{exp(x_{i}W_q(x_{j}W_k)^T/d)}{\sum\limits_{t}exp(x_{i}W_q(x_{t}W_k)^T/d)}
    \end{array}.
\right.
    \label{equ:sa2}
\end{equation} where $\mathcal{A}_i$ represents the set of accessible tokens of token $x_i$. 
As $W_q$ and $W_k$ only attribute in modeling the relationship of queries, once $W_v$ is fixed, the space generated by $x$ is therefore limited. Meanwhile, as shown in Fig.~\ref{fig:domain}(a), SA tends to aggregate ``similar'' features from the tokens, which helps the model to learn long-range dependency. However, because of the nature of US images, long-range contrast is also important to be learned, but SA fails to do so.

Convolution operation could be formulated in the same formulation as SA, as shown in Equ.~\ref{equ:conv}:
\begin{equation}
    \begin{array}{l}
         \widetilde{x}_i =  \sum\limits_{j\in \mathcal{A}_i}\lambda_{j}(x_{j}W_v)
    \end{array}.
    \label{equ:conv}
\end{equation}
In this case, if $rank(x_{j}W_v)=d$, the space generated by $x$ is $\mathbb{R}^{d}$. We randomize $1000$ different pairs of $W_q$ and $W_v$, $1000$ different $\lambda$ and fixed $W_v$. The outputs are shown in Fig.~\ref{fig:domain}. As demonstrated in the figure, the outputs of convolution fulfill the space, while the outputs of SA are gathered. 

A recent work \cite{ding2022scaling} also demonstrates that CNNs with super large kernels learn feature representations from the super large receptive field (sub-global), and achieve comparable performance with state-of-the-art transformers with fewer parameters.

As aforementioned, we need a module to gather long-range textures. As SA fails to learn long-range contrast, and a super-large convolution kernel leads to large computational cost, we need to design a light yet effective convolution module for long-range contrast learning, namely Long-range Contrast Learning Module (LCLM).

\begin{figure}[htbp]
    \centering
    \includegraphics[width=0.4\textwidth]{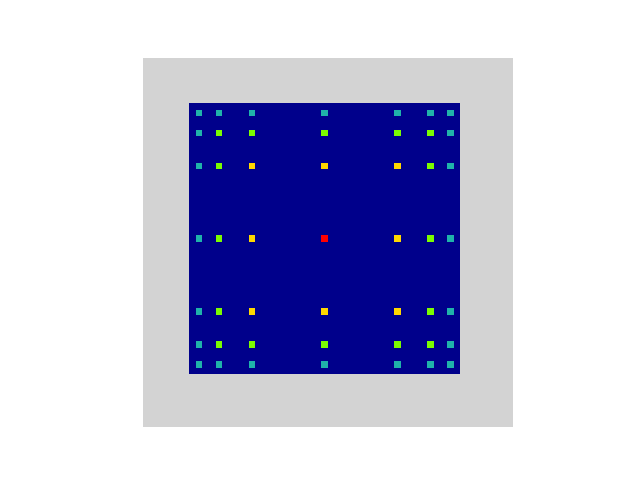}
    \caption{The receptive field of the pixel in red. The pixels painted not in grey are covered by the receptive field.}
    \label{fig:SGM}
\end{figure}

\paragraph{\bf Long-range contrast learning with dilation convolution} To cover long-range contrast with fewer parameters, a simple way is to utilize dilation convolution. In this work, we adopt 3 cascaded dilation convolution layers with dilation \{3,5,11\}. Along with one basic convolution layer which appears at the end of the MobileViT block, LCLM covers up to $41\times 41$ pixels densely. We show the receptive field in Fig.~\ref{fig:SGM}. Each convolution layer is followed by a normalization layer and an activation layer. A depth-wise LCLM only needs $3\times 3\times 3 \times d$ parameters, which is even smaller than a standard convolution layer ($3\times 3\times d \times d$, $d>>3$).

\section{Experiments}

\subsection{Realistic US Simulation from CT Volume}
It is difficult to provide quantitative evaluation criteria for image generation-related works. Qualitatively, the advantages of this paper over deep learning-based US image generation are patient-specific and more accurate lesion boundaries, and when compared with previous papers based on image generation, we make the system produce more realistic images by simulating squeezing and scattering in soft tissues and bones.
Nonetheless, the quantitative evaluation is also of significance, and since the quality of image generation is hard to be evaluated quantitatively, we evaluate the quality of image generation from the purpose of the work.
Except for training doctors to do US examinations, the most important uses of the US image generation algorithms are intra-operative registration and providing datasets for deep learning.
Therefore, we put the generated images into various deep learning-based segmentation algorithms to verify the usefulness of this image generation system for the pre-training of neural networks.

\subsection{Deep Learning-Based Segmentation}
\paragraph{\bf Implementation Details} To validate the proposed US simulation method and the lightweight segmentation model, we train the proposed model on synthetic US images and inference on real US images. We set the batch size to 32, utilize the Adam optimizer, and set the learning rate to 1e-4.

\begin{figure}[hbtp]
    \centering
    \includegraphics[width=0.4\textwidth]{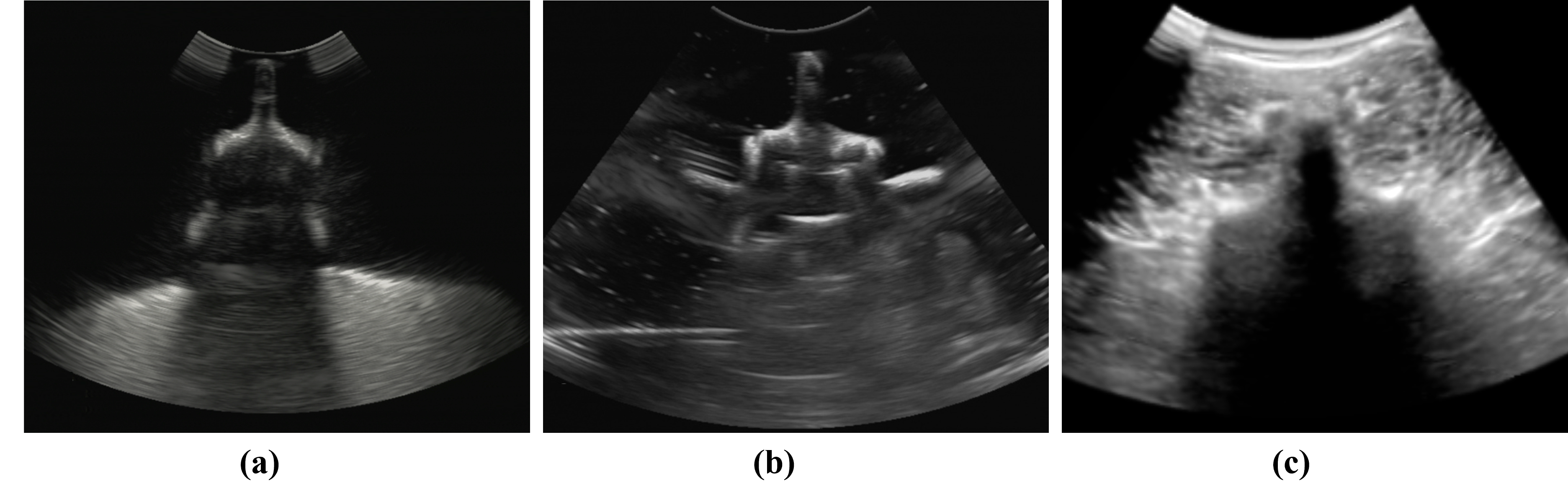}
    \caption{The validation image examples. Figure (a) is the US image of a spinal phantom in gel mimic, Figure (b) is the US image of a 3D-printed spine in water, and Figure (c) is the US image of a clinically human body.}
    \label{fig:val_image_sample}
\end{figure}

\paragraph{\bf Comparison with other models} We first compare the proposed model with other US segmentation models in Tab.~\ref{tab:comp}.
All performances reported in the table are re-implemented by ourselves.
Compared with UNet-based models, the proposed model is much smaller and more effective. Compared with MViT-FPN, the proposed approach achieves much better performance. 

To utilize the model, in this case, the segmentation is to achieve better registration, therefore we remove the cases with obvious segmentation errors to obtain our (selected) result, which is used in the subsequent registration. In this case, the effect of the network trained with generated data can be equivalent to or better than the effect of the SOTA model trained by a large number of real US images with an accurately labeled. Our label is acquired by ct segmentation, which is more challenging for the convergence of lightweight neural networks since he can label bone surfaces which is invisible or ambiguous in some situation in US.

\begin{table}[htbp]
    \centering
    \begin{tabular}{c|c|c|c|c|c}
    \hline
         Method            & Dice  & CD(TP)  &CD(FN)  & \#Parm   & FPS \\
    \hline
         UNet\cite{ronneberger2015u}          
         & 0.574   &  1.085mm &  0.590mm  & 131.8M & 4.13  \\
         MViT-FPN\cite{mehta2021mobilevit}    
         &  0.294  &  4.664mm & 0.824mm  & 20.0M  & 28.1       \\
         CNLUnet\cite{shuvo2021cnl}           
         &  0.329  & 2.868mm  & 1.287mm & 48.6M  & 10.9     \\
         Ours                                 
         & 0.783   &  0.599mm &   1.079mm & 20.0M  & 26.3           \\
         Ours(Selected)                       
         & 0.926   & 0.227mm   & 0.184mm & 20.0M  & 26.3           \\
    \hline
    \end{tabular}
    \vspace{1mm}
    \caption{Comparison with other segmentation models. Note that all performances reported in the table are re-implemented by ourselves.}
    \label{tab:comp}
\end{table}

\paragraph{Ablation study} To demonstrate the effectiveness of synthetic US over initial CT scans, long-range dependency, and long-range contrast, we validate the aforementioned settings and show the results in Tab.~\ref{tab:ablation}. Training the model with CT scans leads to a dramatic IoU decrease, which demonstrates that the proposed method can effectively synthesize US images. Without LCLM, the model fails to learn the long-range contrast of US images and results in a performance decrease. Without MViT, the IoU slightly drops, which represents that the model benefits from long-range dependency, but is less significant compared with long-range contrast.

\begin{table}[htbp]
    \centering
    \begin{tabular}{c|c|c|c|c}
    \hline
         Method        & Baseline    & SUS$\rightarrow$CT    & w/o MViT    & w/o LCLM  \\
         Dice          & 0.783       & 0.040                 & 0.438       & 0.294     \\
         CD(TP)        & 0.599mm     & 6.062mm                 & 0.910mm     & 4.664mm   \\
         CD(FN)        & 1.079mm     & 3.938mm                 & 0.674mm     & 0.824mm   \\
    \hline
    \end{tabular}
    \vspace{1mm}
    \caption{The ablation study of the proposed model. ``SUS'' represents the synthetic US. ``SUS$\rightarrow$CT'', ``w/o MViT'' and ``w/o LCLM'' represent training the model by CT scans, removing MViT from the baseline and removing LCLM from the baseline, respectively.}
    \label{tab:ablation}
\end{table}

\begin{figure}[htbp]
    \centering
    \includegraphics[width=0.4\textwidth]{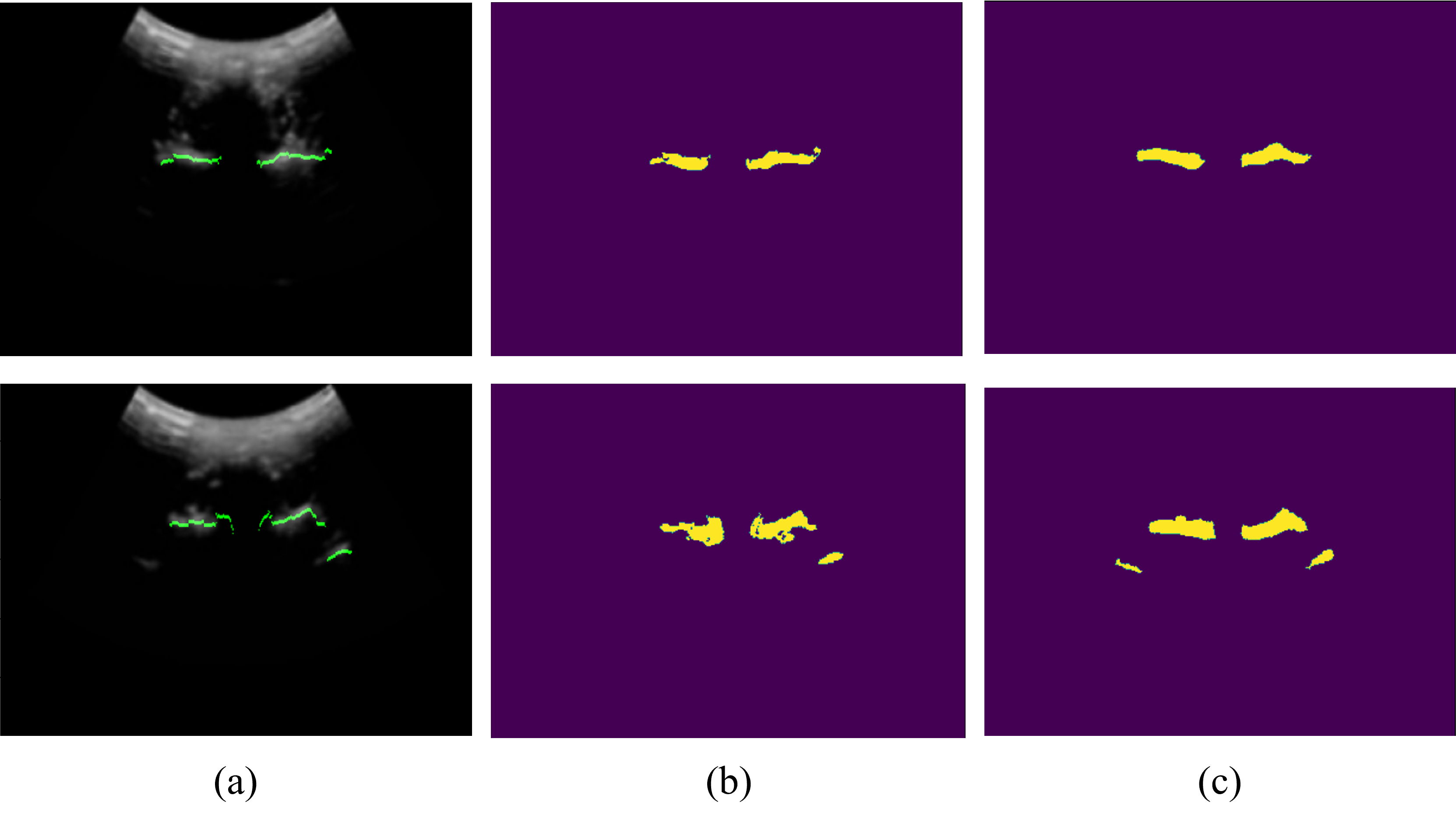}
    \caption{The result of segmentation. Figure(a), (b), and (c) presents the segmented label, the output of the network, and the ground truth respectively.}
    \label{fig:regi_result}
\end{figure}

\subsection{Validation of US-CT Registration}

\paragraph{Implementation Details}

The US simulation system is proposed to better obtain labeled data from multiple poses, patients, and environments, enabling US-based surgical navigation to be accomplished.
In this paper, we validated whether the output of the model trained on synthetic data can align the preoperative CT and the planned trajectory into the intra-operative phase, and the spinal bone surface registration for pedicle screw placement is the instance.

The validation is divided into two parts, the first is to validate the accuracy of the rigid body registration, i.e. translation and rotation.
The second part is to validate whether the pre-operatively planned trajectory with the aforementioned transforms will cause complications for the surgery or not, i.e. whether the intra-operative trajectory will touch vital organs.

The experiments were performed on three phantoms, a spine phantom in water, a 3D-printed human spine in water, and a bovine spine in agar gel.
The registration steps were divided into a coarse alignment and an individual registration for each segment. The coarse registration used all point clouds to calculate the approximate position and orientation, and the final registration used a de-noised Iterative Closest Point(ICP) method.
The results of the registration are shown in the table below.

\begin{figure}[htbp]
    \centering
    \includegraphics[width=0.4\textwidth]{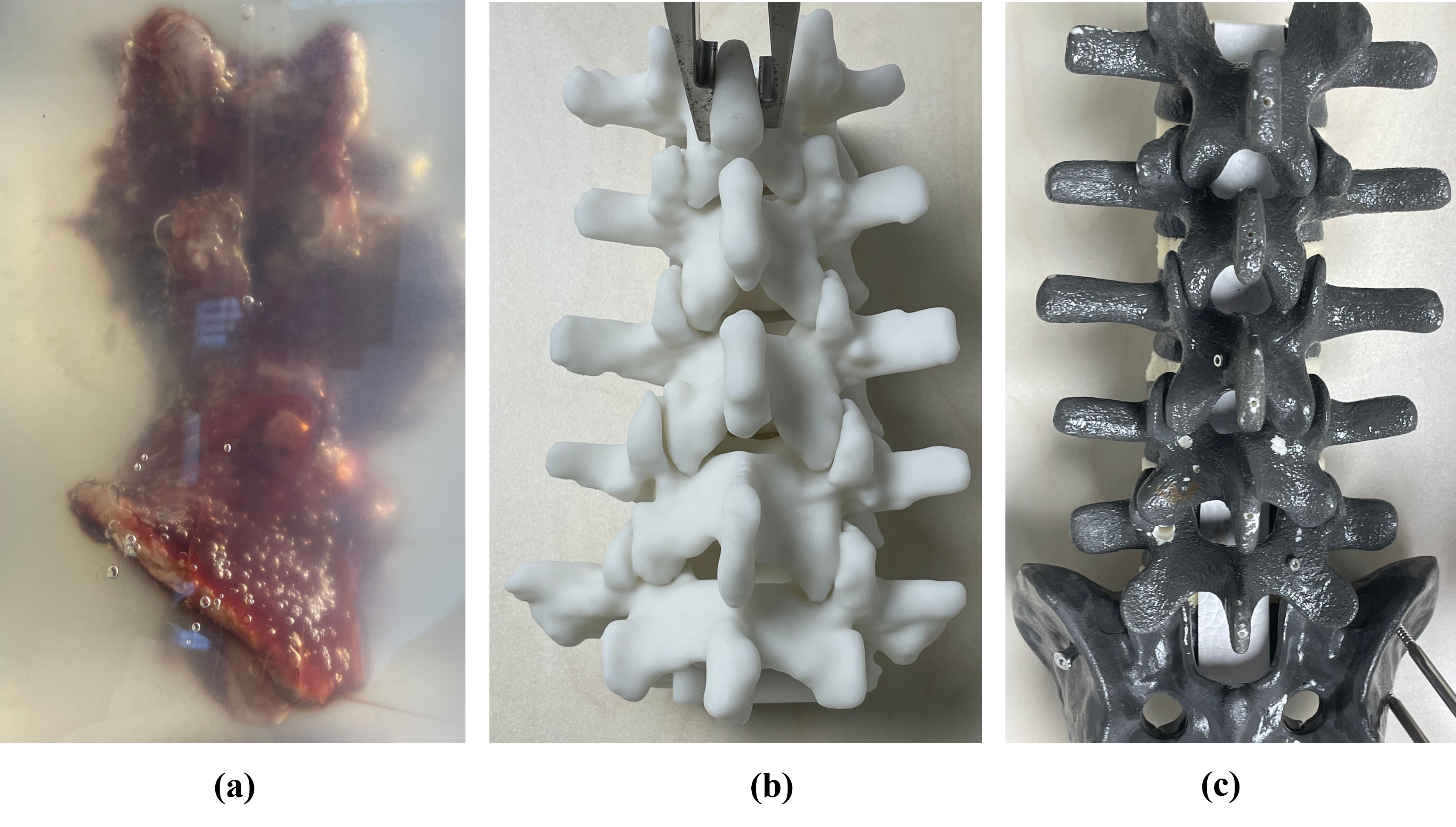}
    \caption{The phantom model for registration validation.}
    \label{fig:registration_phantom}
\end{figure}

\begin{figure}[htbp]
    \centering
    \includegraphics[width=0.4\textwidth]{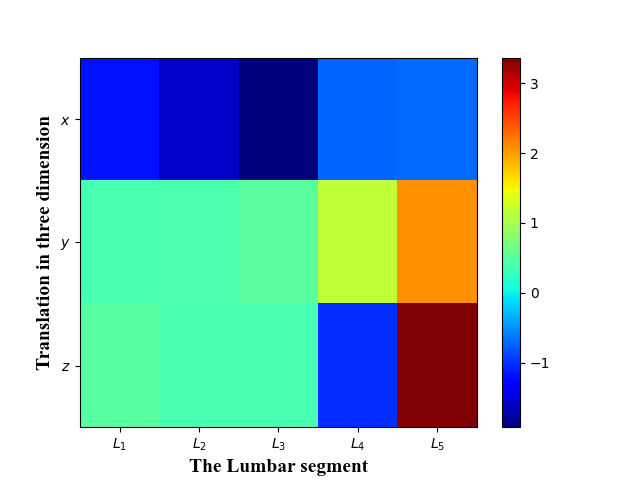}
    \caption{The MSE loss of registration in three dimensions.}
    \label{fig:regi_acc}
\end{figure}

\begin{figure}[htbp]
    \centering
    \includegraphics[width=0.4\textwidth]{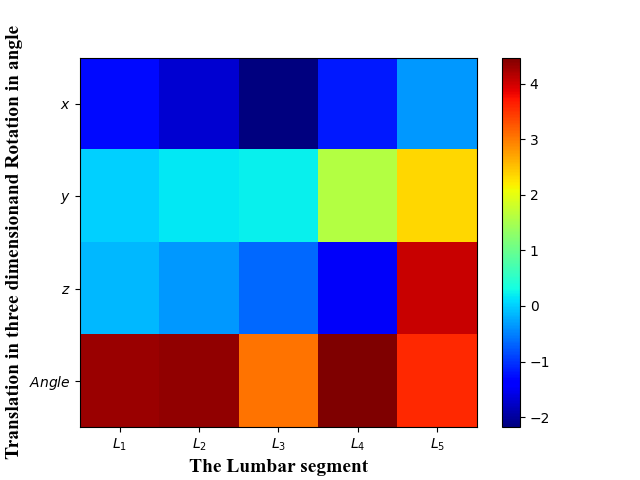}
    \caption{The Translation and Rotation of pedicle screw with registration matrix. The rotation is presented as an angle in degrees.}
    \label{fig:regi_rt}
\end{figure}

\paragraph{Validation of intra-operative model registration}

In the registration phase, the model inputs the US image outputs the segmentation label, and extracts the upper contour of its maximum connectivity, hence the point cloud is created with the calibration matrix of the US probe. The calibration of the US is done by two cross-wire phantom \cite{2013Improving}, with 0.97mm MSE(min square loss) loss.
Since the coordinate system of each segment does not coincide, the rotation evaluation in angular will be represented by the rotation of pedicle screws in the next part.
To evaluate the error of registration, the point-to-point MSE is used, which is also the objective function of the ICP algorithm.
The result of registration is shown in Fig.~\ref{fig:regi_result}, which shows the ground truth in the intra-operative space(white model) and the results of registration with US (green model).
In summary, the segmentation algorithm trained by the generated US image works in the registration with the error $0.133 \sim 3.366$mm. The maximum appears in the z-axis of L5, which is around 3 mm, which is probably caused by the shape of L5 being more different from the shape of other lumbar.

\paragraph{Validation of Pedicle Screw Placement Feasibility}

\begin{figure}[htbp]
    \centering
    \includegraphics[width=0.4\textwidth]{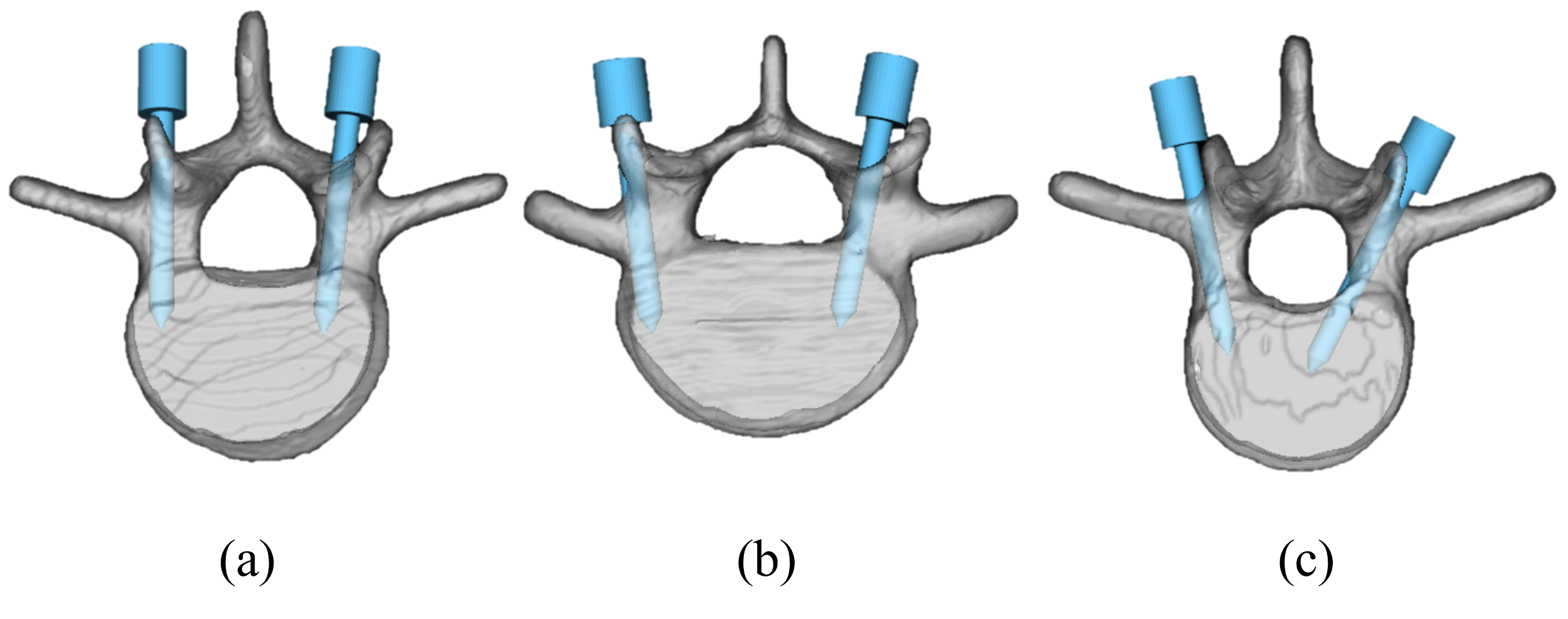}
    \caption{The visualization of registration-based screw placement.  The figure shows the posture of the screws with standard length and diameter which is planned on preoperative CT by an experienced surgeon, compared to the intra-operative spine ground truth by the registration matrix acquired by image segmentation. Cases derived from the standard spine phantom, which is (c) in the figure ~\ref{fig:registration_phantom}.}
    \label{fig:screw_placement}
\end{figure}

To verify the registration effectiveness of the system in the target tasks, we compared the translation and rotation of the tip of the screw with the US point cloud registration and ground truth using an expert-planned intra-operative plan for pedicle screw placement. The rotation is determined by the angle between two pedicle axis vectors in 3D space which is expressed in degrees. 
As shown in Fig. \ref{fig:regi_rt}, the error in translation for the tip of the screw ranges from 0.027 mm to 4.031 mm, with the maximum value still occurring on the z-axis of L5, which is consistent with the above assessment.
The error in the rotation is between 3.05 degrees and 4.47 degrees, therefore, it can be concluded that the pre-trained model with generated US image can perform the alignment with small errors. The results of the pedicle screw placement are shown in figure \ref{fig:screw_placement}, which shows this registration does not put the patient at risk of complications.

\begin{figure}[htbp]
    \centering
    \includegraphics[width=0.4\textwidth]{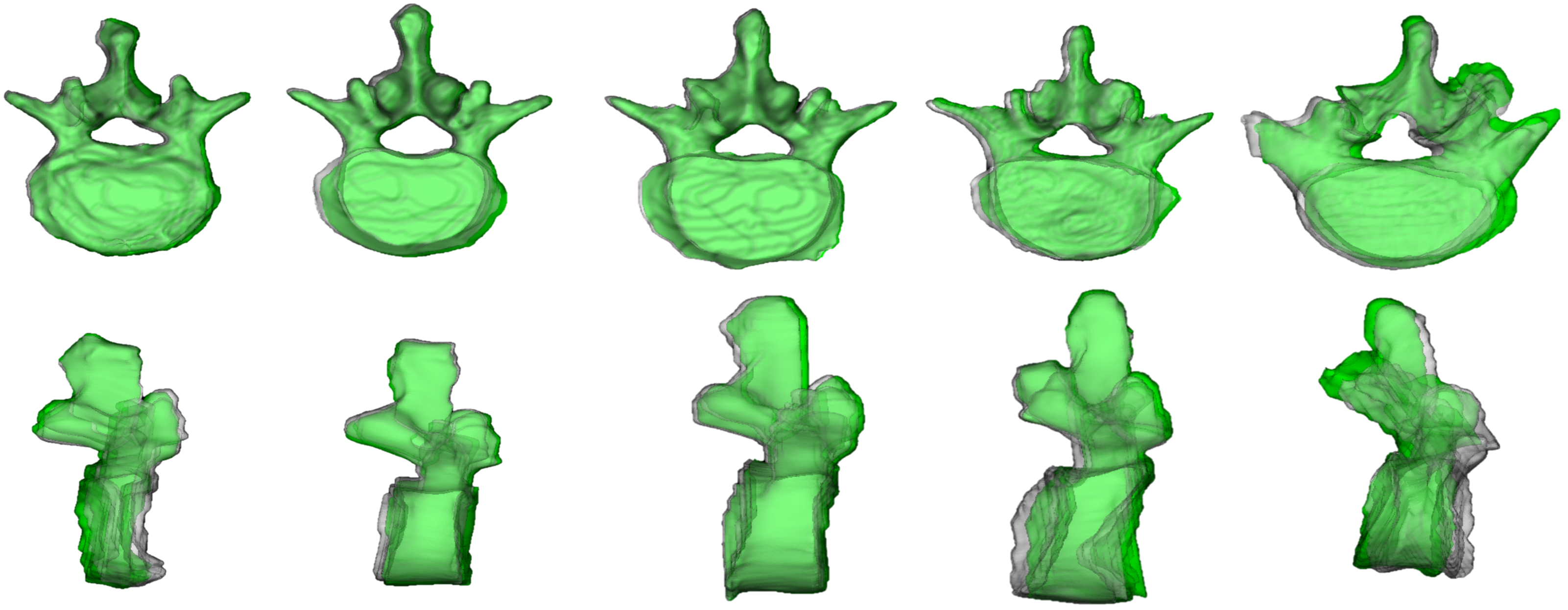}
    \caption{The visualization of registration. The green part of the figure is the registration result by US segmentation, and the white part is the ground truth. Cases derived from 3d printed patient spine, which is (b) in the figure ~\ref{fig:registration_phantom}.}
    \label{fig:regi_result}
\end{figure}

\section{Conclusion}
In this paper, we propose a US image simulation method based on CT images and acoustic properties that automatically generates a US simulation environment based on a specific patient CT volume, which allows US-based deep learning and reinforcement learning algorithms to be trained and validated. In addition, we propose a lightweight vision transformer for segmenting US images which is a network structure with better segmentation accuracy and modal generalization capabilities.

Experiments show that the model trained using US images generated by the realistic US simulation from the CT system achieves higher segmentation accuracy and has the capability to perform the intraoperative registration process without complications compared to the model trained with CT images. Compared to other segmentation algorithms, the proposed transformer has real-time computational efficiency, better segmentation accuracy, and generalization capability, which is particularly important in US surgical applications.

\bibliographystyle{IEEEtran}
\bibliography{IEEERef}

\end{document}